\documentclass[nocopyright]{cimento}
\usepackage{physics}
\usepackage{tikz}
\usepackage{slashed}
\usepackage{pgfplots}
\usepackage{mathtools}
\usepackage{siunitx}
\usepackage{physics}
\usepackage[compat=1.1.0]{tikz-feynman}
\usepackage[T1]{fontenc}
\usepackage{multirow}
\usepackage{booktabs}
\usepackage{cleveref}
\usepackage{tabularray}
\usepackage{nicematrix}
\usepackage{adjustbox}
\usepackage{tabularx}
\newcommand{\ptlike}{0}

\newcommand{\ifi}{\text{IFI}}

\newcommand{\disp}{\text{FsQED}}

\newcommand{\lo}{\text{LO}}
\newcommand{\Amp}{\mathcal{M}}
\title{On the inclusion of the pion form factor in $e^+e^-\to \pi^+\pi^-$ beyond leading order}

\author{Francesco P. Ucci\from{unipv}\from{infn-pv} }
\instlist{
\inst{unipv}{Dipartimento di Fisica "Alessandro Volta", Universit\`a di Pavia, Via A. Bassi 6, 27100 Pavia, Italy}
\inst{infn-pv}{INFN, Sezione di Pavia, Via A. Bassi 6, 27100 Pavia, Italy}}
\begin{document}
\maketitle
\vspace{-1cm}
\begin{abstract}
The pion form factor plays a crucial role in the determination of the contribution of the hadronic vacuum polarisation to the muon anomalous magnetic moment. In order to measure this quantity, energy-scan experiments rely on Monte Carlo generator to simulate the $e^+e^- \to \pi^+\pi^-(\gamma)$ process. For the theoretical accuracy to match the experimental precision, next-to-leading order calculations and the resummation of multiple photon emissions are needed. In this context, the inclusion of the pion form factor beyond the leading order approximation is crucial to reproduce some observables, like the pion charge asymmetry. We present the impact of the inclusion of the pion form factor in loop diagrams with three approaches and the interplay with radiative corrections.
\end{abstract}
\vspace{-1.4cm}
\section{Introduction}
Most of the theoretical uncertainty of the muon $g-2$ anomaly~\cite{Jeger} comes from the hadronic vacuum polarisation contribution, with more than the 70\% of it given by the $\pi\pi$ channel.
In the Standard Model, such effects can be computed with two alternative approaches: the \textit{ab-initio} lattice calculation~\cite{WP25}, which is in perfect agreement with the latest FNAL experimental results, or the dispersive approach, which relies on the measure of the $e^+e^-\to \textit{hadrons}$ cross section, as compiled in~\cite{WP20}. The latter is discrepant at the level of $5.7\sigma$ with the experimental determination but latest CMD-3 measurement~\cite{CMD3-meas} of the pion form factor ($\pi$FF) via energy scan, shifts the theoretical prediction at $1\sigma$ level of agreement with the measured $a_\mu=(g-2)_\mu/2$. \\
In this puzzling scenario, attention has been put on the impact of radiative corrections~\cite{Aliberti:2024fpq}, especially in the $\pi\pi$ channel that amounts to the measurement of the pion electromagnetic form factor. For example, out of the 0.7\% systematic error of the CMD-3 result~\cite{CMD3}, the $0.3\%$ is given by differences in higher order effects computed by the Monte Carlo (MC) generators used in the analysis.
Moreover, it has been shown in the literature that the compositeness has to be taken into account in next-to-leading-order (NLO) calculations~\cite{GVMD,FsQED}. It is therefore necessary to go beyond the usual factorised scalar QED (F$\times$sQED) approximation, with approaches based alternatively on the generalised vector meson dominance (GVMD) model or dispersive relations (FsQED).\\
\section{NLOPS accuracy in factorised SQED}
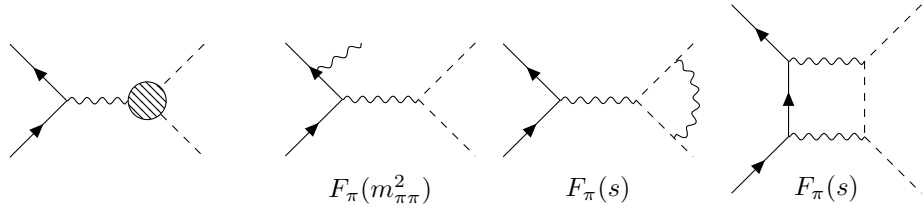
\begin{figure}
\begin{equation*}
\begin{gathered}
     \begin{tikzpicture}
  \begin{feynman}[small][baseline=(a.base)]
    \vertex (a) ;
    \vertex[right=0.8cm of a,style=blob] (b) {};
    \vertex[above left=0.75cm and 0.75cm of a] (c);
    \vertex[below left=0.75cm and 0.75cm of a] (d);
    \vertex[above right=0.75cm and 0.75cm of b] (e);
    \vertex[below right=0.75cm and 0.75cm of b] (f);
    \vertex[above left =0.5 and 0.5cm of a] (g);
    \vertex[below left =0.5 and 0.5cm of a] (h);
    \vertex[above left=0.35cm and 0.35cm of a](i);
    \vertex[above right=0.4cm and 0.6cm of i](j);
    \diagram* {
      (a) -- [photon] (b),
      (d) -- [fermion] (a),
      (a) -- [fermion] (c),
      (e) -- [scalar] (b) --[scalar] (f),
    };
  \end{feynman}
\end{tikzpicture}
\end{gathered}    
\qquad\quad
\begin{gathered}
  \vspace{-20pt}
     \begin{tikzpicture}
  \begin{feynman}[small]
    \vertex (a) ;
    \vertex[right=1cm of a] (b);
    \vertex[right=0.5cm of a, label={[yshift=-25pt]below:$ F_\pi(m_{\pi\pi}^2)$}] (k);
    \vertex[above left=0.75cm and 0.75cm of a] (c);
    \vertex[below left=0.75cm and 0.75cm of a] (d);
    \vertex[above right=0.75cm and 0.75cm of b] (e);
    \vertex[below right=0.75cm and 0.75cm of b] (f);
    \vertex[above left =0.5 and 0.5cm of a] (g);
    \vertex[below left =0.5 and 0.5cm of a] (h);
    \vertex[above left=0.35cm and 0.35cm of a](i);
    \vertex[above right=0.4cm and 0.6cm of i](j);
    \diagram* {
      (a) -- [photon] (b),
      (d) -- [fermion] (a),
      (a) -- [fermion] (c),
      (e) -- [scalar] (b) --[scalar] (f),
      (i) -- [photon] (j),
    };
  \end{feynman}
\end{tikzpicture}
\end{gathered}    
\quad
\begin{gathered}
  \vspace{-20pt}
  \begin{tikzpicture}
  \begin{feynman}[small]
    \vertex (a);
    \vertex[right=1cm of a] (b);
      \vertex[right=0.5cm of a, label={[yshift=-25pt]below:$F_\pi(s)$}] (k);
    \vertex[above left=0.75cm and 0.75cm of a] (c);
    \vertex[below left=0.75cm and 0.75cm of a] (d);
    \vertex[above right=0.75cm and 0.75cm of b] (e);
    \vertex[below right=0.75cm and 0.75cm of b] (f);
    \vertex[above right =0.5 and 0.5cm of b] (g);
    \vertex[below right =0.5 and 0.5cm of b] (h);
    \diagram* {
      (a) -- [photon] (b),
      (d) -- [fermion] (a),
      (a) -- [fermion] (c),
      (e) -- [scalar] (b) --[scalar] (f),
      (g) --[photon,half left, looseness=1,] (h),
    };
  \end{feynman}
\end{tikzpicture}
\end{gathered}
\quad
\begin{gathered}
\vspace{-5pt}
     \begin{tikzpicture}
  \begin{feynman}[small]
    \vertex (a);
    \vertex[right=1cm of a] (b); 
    \vertex[below=1cm of a] (i);
    \vertex[below=1cm of b] (j);
        \vertex[right=0.5cm of i, label={[yshift=-11pt]below:$ F_\pi(s)$}] (k);
    \vertex[above left=0.75cm and 0.75cm of a] (c);
    \vertex[below left=0.75cm and 0.75cm of i] (d);
    \vertex[above right=0.75cm and 0.75cm of b] (e);
    \vertex[below right=0.75cm and 0.75cm of j] (f);
    \diagram* {
      (a) -- [photon] (b),
      (d) -- [fermion] (i) -- [fermion] (a) -- [fermion] (c),
      (e) -- [scalar] (b) --[scalar] (j) -- [scalar] (f),
      (i) -- [photon] (j),
    };
  \end{feynman}
\end{tikzpicture}      
\end{gathered}
\end{equation*}
\caption{Feynman diagrams for the $e^+e^-\to\pi^+\pi^-$ process. From left to right: tree level diagram, example of real ISR, virtual FSR vertex correction and virtual IFI.}
\label{fig:DIAGS}
\end{figure}

The pion form factor $F_\pi(q^2)$ is defined as the expectation value of the light quark electromagnetic current $j_\text{em}^\mu$ between pion states
\(    \langle \pi^\pm(p') | j_\mathrm{em}^\mu | \pi^\pm(p) \rangle =\pm (p'+p)^\mu F_\pi\left((p'-p)^2\right)
\),
accounting for the charge distribution of the pion. In order to fulfil the Ward identities, the condition $F_\pi(0)=1$ holds, meaning that real photons do not resolve the pion compositeness. Therefore, the Born approximation for the differential cross section of the process $e^+e^- \to \gamma^* \to \pi^+\pi^-$, shown in the first diagram of fig.~\ref{fig:DIAGS}, is given by 
\begin{equation}\label{eq:born}
      \dv{\sigma_{\rm{LO}}}{\cos\vartheta}=\frac{\alpha^2\pi}{4s\beta_e}\beta_\pi^3(1-\beta_e^2\cos^2\vartheta) |F_\pi(s)|^2 \,,
\end{equation}
where the velocity of the particle is defined as $\beta_i=\sqrt{1-4m_i^2/s}$, with $s$ the centre of mass energy. 
At NLO, the cross section receives $\order{\alpha}$ virtual photonic corrections, contributing to the 2$\,\to\,$2 process, and the emission of a real photon that gives the $e^+e^-\to\pi^+\pi^-\gamma$ signature; vacuum polarisation effects are already included in the definition of the form factor. The differential cross section in the scattering angle can be decomposed in initial-state radiation (ISR), final (FSR) and intitial-final interference (FSR) contributions to the soft and virtual cross sections, as depicted in fig.~\ref{fig:DIAGS}, plus the hard emission, as follows
\begin{equation}\label{EQ:crossecsplitting}
    \dv{\sigma_\text{NLO}}{\cos\theta}= \dv{\sigma_\text{LO}}{\cos\theta} \left(1 + \delta_{SV}^\text{ISR} + \delta_{SV}^\text{FSR} +\delta_{SV}^\text{IFI}\right) + \dv{\sigma_H}{\cos\theta}\, .
\end{equation}
Conversely to the Born approximation, the IFI soft plus virtual correction and the hard photon emission are not even under the exchange $\cos\theta\to-\cos\theta$~\cite{ArbuREV}, generating a non-vanishing forward-backward asymmetry, defined as
\begin{equation}\label{eq:FBasym}
    A_{\rm FB}\left(\sqrt{s}\right)=\frac{\sigma_{\rm F}-\sigma_{\rm B}}{\sigma_{\rm F}+\sigma_{\rm B}} \,.
\end{equation}
In F$\times$sQED, each amplitude is multiplied by  evaluated at the virtuality of the photon propagator, or at $q^2=s$ for virtual box diagrams (see fig.~\ref{fig:DIAGS}). This prescription ensures the cancellation of infrared (IR) divergences between the virtual and soft diagrams, which in the case of IFI diagrams arise when the virtuality of one of the two photons goes to zero $q_1^2\to s, \, \,q_2^2\to0$. However, the multiplication of IFI diagrams by $F_\pi(s)$ does not take into account the phase-space region where both photons virtualities are hard. \\
On top of the fixed-order calculations, the exclusive generation of additional photons in treated by means of a Parton Shower algorithm~\cite{PS_BY}.
In the PS approach, no further complication on top of the fixed-order calculation arises due to the inclusion of $F_\pi(q^2)$.
\section{Inserting $F_\pi(q^2)$ in the loop}
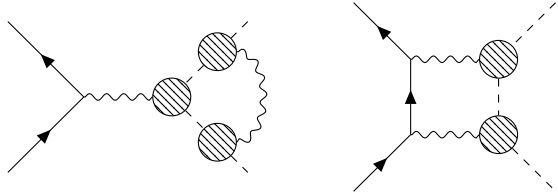
\begin{figure}
\begin{equation*}
\begin{gathered}
  \begin{tikzpicture}
\begin{feynman}[small]
    \vertex (a);
    \vertex[right=0.9cm of a,style=blob] (b) {};
    \vertex[above left=1cm and 1cm of a] (c);
    \vertex[below left=1cm and 1cm of a] (d);
    \vertex[above right=1cm and 1cm of b] (e);
    \vertex[below right=1cm and 1cm of b] (f);
    \vertex[above right =0.6and 0.6 of b,style=blob] (g) {};
    \vertex[below right =0.6and 0.6cm of b,style=blob] (h) {};
    \diagram* {
      (a) -- [photon] (b),
      (d) -- [fermion] (a),
      (a) -- [fermion] (c),
      (e) -- [scalar] (g) --[scalar] (b),
      (b) -- [scalar] (h) --[scalar] (f),
      (g) --[photon,half left, looseness=1,] (h),
    };
\end{feynman}
\end{tikzpicture}
\end{gathered}
\quad\qquad
\begin{gathered}
     \begin{tikzpicture}
  \begin{feynman}[small]
    \vertex (a);
    \vertex[right=0.9cm of a,style=blob] (b) {}; 
    \vertex[below=1cm of a] (i);
    \vertex[below=1cm of b,style=blob] (j) {};
    \vertex[above left=0.75cm and 0.75cm of a] (c);
    \vertex[below left=0.75cm and 0.75cm of i] (d);
    \vertex[above right=0.75cm and 0.75cm of b] (e);
    \vertex[below right=0.75cm and 0.75cm of j] (f);
    \diagram* {
      (a) -- [photon] (b),
      (d) -- [fermion] (i) -- [fermion] (a) -- [fermion] (c),
      (e) -- [scalar] (b) --[scalar] (j) -- [scalar] (f),
      (i) -- [photon] (j),
    };
  \end{feynman}
\end{tikzpicture}      
\end{gathered}
\end{equation*}
\caption{Virtual FSR and ISR Feynman diagrams, where the blob amounts to the insertion of $F_\pi(q^2)$ inside the loop integral with the proper virtuality.}
\label{fig:DIAGS_FF}
\end{figure}

In order to overcome the factorised approach limitations by including $F_\pi(q^2)$ inside the loop calculation, one needs an analytic representation for the $\pi$FF that allows to compute loop integrals by means of standard techniques. Diagrammatically, this amounts to insert the form factor in all vertices by multiplying the sQED Feynman rules by $F_\pi(q^2)$ considering the photon virtuality $q^2$, as depicted in fig.~\ref{fig:DIAGS_FF}.
As the FSR is already included in the measurement of the form factor, we limit our discussion to the IFI\footnote{For completeness of the theoretical description, as well as to estimate the effect for experimental cross-checks, also the FSR diagrams are included in~\cite{Budassi2024}.}, given by box diagrams like in the right panel of fig.~\ref{fig:DIAGS_FF}, while the ISR is not modified w.r.t. sQED. Therefore, we define the virtual correction as the interference of the LO amplitude in sQED and the virtual one in the GVMD and FsQED approaches 
\begin{equation}
\delta_{V,\text{FF}}^\text{IFI}(\lambda) = \frac{ 2\Re{F_\pi(s)^*\Amp_{\lo,\ptlike}^\dagger \,\Amp_{V,\text{FF}}^\ifi}}{\left|F_\pi(s)\right|^2|\Amp_{\text{LO},0}|^2},\quad \text{FF}=\text{GVMD,\,FsQED}.
\end{equation}
The above quantity can be written as sums or integrals of the point-like correction with massive virtual photons $ \bar{\delta}_{V}^{\ifi}(s',s'')$, whose analytical expression is given in~\cite{Budassi2024},
where $m_\gamma^2=s',s''$ is the photon effective mass squared.
\subsection{Generalised vector meson dominance}
Proposed by~\cite{GVMD}, the GVMD approach is based on approximately writing the pion FF as a sum of Breit-Wigner (BW) functions, stemming for the physical resonances that decay in the $\pi\pi$ channel, 
\begin{align}
F_\pi(q^2) = \sum_{v=1}^{n_r} F^\text{BW}_{\pi,v}(q^2)  = \frac{1}{c_t} \sum_{v=1}^{n_r} c_v \frac{\Lambda_v^2}{\Lambda_v^2 - q^2} \:,
\label{eq:bwsum}
\end{align}
where  $\Lambda_v^2 = m_v^2 - i m_v \Gamma_v$ is the complex mass of each resonance, $c_v = |c_v|e^{i\phi_v}$ a complex coefficient and $c_t=\sum_vc_v$ ensures the condition $F_\pi(0)=1$. This model amounts to multiplying each photon propagator by a resonant one before performing the loop integration. However, due to a simple propagator identity, it can be rewritten as the sum of contributions involving in which one or two virtual photons have a resonance mass $\Lambda_v$, so that the correction becomes
\begin{equation}\begin{aligned}\label{eq:deltagrossa_gvmd}
\delta_{V,\text{GVMD}}^\text{IFI}=\hspace{-5pt}\sum_{v=1}^{n_r}\sum_{w=1}^{n_r}\text{Re}\biggl\{\frac{c_v  c_w}{c_t^2 F_\pi(s)}\Bigl[&
\Bar\delta_{V}^\text{\hspace{1pt}IFI}(\lambda^2,\lambda^2)
-\Bar\delta_{V}^\text{\hspace{1pt}IFI}(\Lambda_v^2,\lambda^2)\\
&
-\Bar\delta_{V}^\text{\hspace{1pt}IFI}(\lambda^2,\Lambda_w^2)
+\Bar\delta_{V}^\text{\hspace{1pt}IFI}(\Lambda_v^2,\Lambda_w^2)\Bigr] \biggr\}\,,
\end{aligned}
\end{equation}
where $\lambda^2$ is an infinitesimal photon mass that regularises IR divergences. In the soft limit, the IR divergence exactly cancels with the real soft emission, as the above equation reduces to $\Bar{\delta}_V^\ifi(\lambda^2,\lambda^2)|_\text{IR}$, which has the same divergent structure of the point-like amplitude. Despite the GVMD model being a good description of asymmetry data, as shown in Ref.~\cite{CMD3-meas}, Breit-Wigner functions limit the goodness of the data fit, other than introducing a sub-threshold imaginary part that does not comply with the unitarity of the scattering matrix.
\subsection{Dispersive approach} The pion form factor is an analytic function on the whole complex plane, except for a branch cut at $s\geq 4m_\pi^2$ corresponding to the physical threshold. Relying on its analytical properties, one can write the pion form factor as a once-subtracted dispersion relation~\cite{FsQED} decorated by a sum rule 
\begin{equation}
F_\pi(q^2) = 1 + \frac{q^2}{\pi} \int_{4m_\pi^2}^\infty \frac{\dd s'}{s'}\frac{\Im F_\pi(s')}{s'-q^2-i\varepsilon'}\,\qquad \quad\frac{1}{\pi} \int_{4m_\pi^2}^\infty \frac{\dd s'}{s'} \Im F_\pi(s')= 1\, .
\label{eq:sumrule}
\end{equation}
The subtraction at $s=0$, along with the sum rule, enforces the normalisation $F_\pi(0)=1$, as well ensures the convergence for large $s$. Practically, the integral is extended by a certain energy cutoff, depending on the data. When the dispersion relation is inserted twice in the box amplitudes the factor $F_\pi(q^2)/q^2$ amounts to an effective massive propagator whose mass is integrated upon, giving rise to double polar, polar-dispersive and double dispersive terms, as
\begin{equation}
    \begin{aligned}
\delta_{V,\disp}^\ifi=\frac{1}{|F_\pi(s)|^2}\Re\biggl\{F^*_\pi(s)\Bigl[&\Bar{\delta}^\ifi_V(\lambda^2,\lambda^2) 
-\frac{2}{\pi} \int_{4 m_\pi^2}^{\infty} \frac{\dd s'}{s'} \Im F_\pi(s') \Bar{\delta}^\ifi_V(s',\lambda^2)\\
     &+  \frac{1}{\pi^2}\int_{\Omega_\infty}
\frac{\dd s'}{s'}  \frac{\dd s''}{s''}\Im F(s') \Im F(s'')\Bar{\delta}^\ifi_V(s', s'')\Bigr]\biggr\}\,.
    \end{aligned} \label{eq:realdisp}
\end{equation}
In this representation, since the loop and dispersive integrals have imaginary parts, IR divergence is spread between the real and imaginary part of the massive-photon kernels. In order to retrieve it, we carefully regularise the integrals by numerically subtracting the term originating the IR divergence and adding it analytically. The MC integration is performed by generating one value of the pair $(s',s'')$ for each phasespace point.
\section{Phenomenological results}
In order to fix the parameters of the form factor, we use a combination of energy scan data: the CMD-3~\cite{CMD3-meas} from the threshold to $\sqrt{s}\le 1.1~\rm{GeV}$, the CMD-2 measurements~\cite{CMD2} in the range $1.1~\rm{GeV}<\sqrt{s}\leq 1.35~\rm{GeV}$ and the DM-2 data~\cite{DM2} up to $\sqrt{s}=2.12~\rm{GeV}$, considering the $v=\rho,\rho',\rho'',\omega,\phi$ meson resonances.
A realistic representation of the form factor, as done in experimental fits, is given by a combination of Breit-Wigner and Gounaris-Sakurai (GS) functions
\begin{equation}
\label{eq:FF_GS}
F_\pi^\textrm{GS}(q^2) =\frac{1}{c_t} \Bigg[ \Biggl(1+ \sum_{v=\omega,\phi}c_v\,\frac{q^2}{m_v^2}\textrm{BW}_v\Biggr) \textrm{BW}_\rho^\textrm{GS}+ c_{\rho'}\, \textrm{BW}_{\rho'}^\textrm{GS}+ c_{\rho''}\, \textrm{BW}_{\rho''}^\textrm{GS} \Bigg] \,,
\end{equation}
where $c_t=1+c_{\rho'}+c_{\rho''}$, which resembles the analytic properties that the $\pi$FF should have from first principles. We use the this form factor for the F$\times$sQED and FsQED approaches, while the GVMD approach requires the form factor to be written in the form of Eq.~\eqref{eq:bwsum}. In the left plot of fig.~\ref{fig:scan+FF}, we show that the difference of the absolute value squared of the form factor between the GS and BW parametrisation, with the numerical values given in tab. 2 of Ref.~\cite{Budassi2024}, can be as large as $4\%$.\\
The numerical results are obtained via the latest version of \textsc{BabaYaga@NLO}~\cite{Budassi2024}, in which the $e^+e^-\to\pi^+\pi^-$ process is implemented as discussed above, using a set of cuts on the phase space: the $\pi^\pm$ spacial momenta are taken $|\vec p^\pm|>0.45 \sqrt{s}/2$, the average polar angle $\theta_\text{avg}=1/2(\pi-\theta^++\theta^-)\in[1,\pi-1]$ and the polar and azimuthal acollinearity are taken as $\delta\theta<0.25,\delta\phi<0.15$.
\begin{figure}
    \centering
    \includegraphics[width=0.8\linewidth]{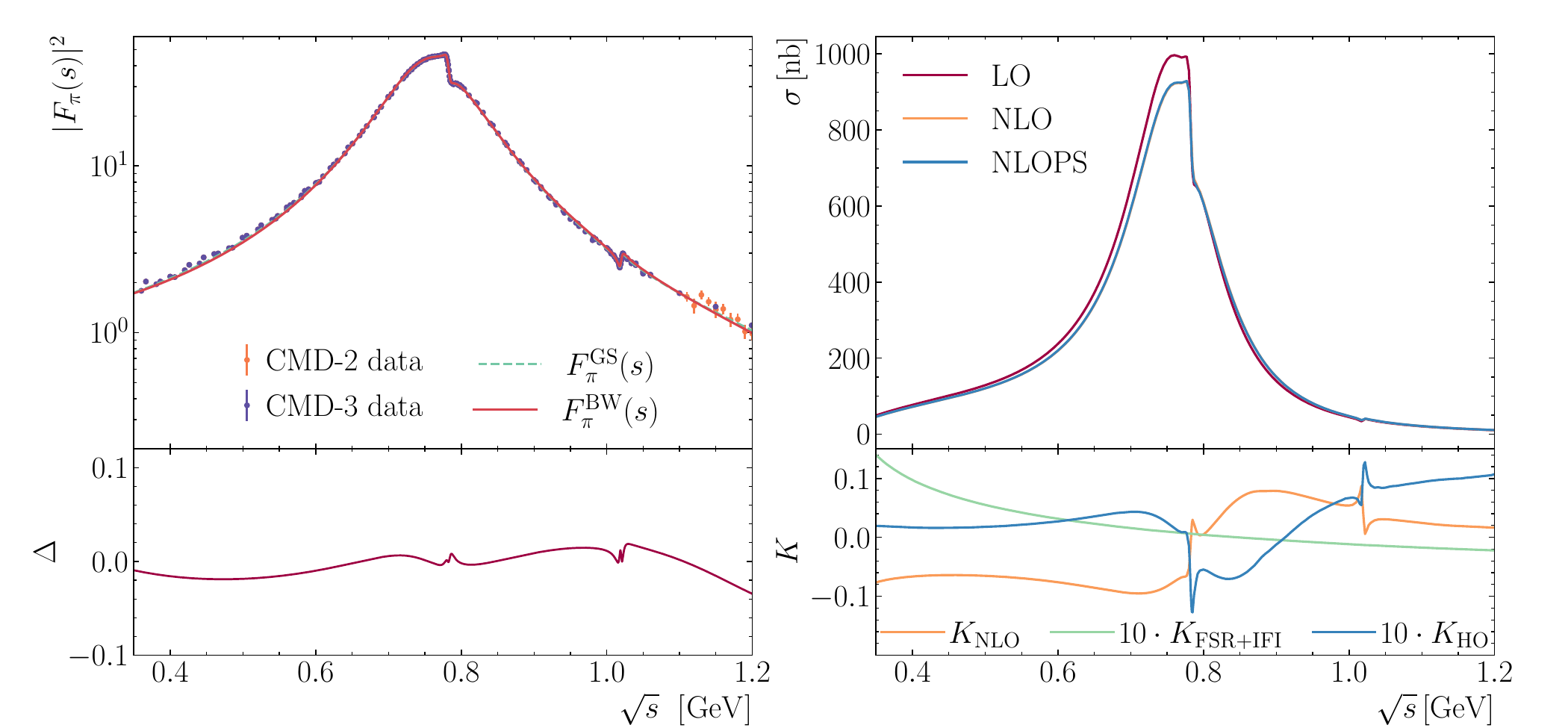}
    \caption{The left plot shows the absolute value squared of the pion form factor, fitted with the GS and BW parametrisations, the bottom panel displays $\Delta = |F_\pi^\text{GS}|/|F_\pi^\text{BW}|-1$. To the right, the integrated cross section as a function of $\sqrt{s}$ at three levels of accuracy.}
    \label{fig:scan+FF}
\end{figure}
In the right plot of fig.~\ref{fig:scan+FF} we show the integrated cross section from the $4m_\pi^2$ up to $1.2~\rm{GeV}$ at LO, NLO and at NLOPS. The ratio $K_\text{NLO}=(\sigma_\text{NLO}-\sigma_\text{LO})/\sigma_\text{LO}$ shows that the effect of $\order{\alpha}$ corrections ranges from $-8\%$ to $+8\%$, the change of sign being a typical feature of resonances. The green line shows the effect of the FSR, since the IFI vanishes, defined as $K_\text{FSR+IFI}=(\sigma_\text{NLO}-\sigma_\text{ISR})/\sigma_\text{LO}$. As expected, the effect of the radiation from pions is an order magnitude smaller with respect to the ISR, which has a logarithmic enhancement $\log(s/m_e^2)\simeq 12 \log(s/m_\pi^2)$. The effect of higher order (HO) photon emission $K_\text{HO}=(\sigma_\text{NLOPS}-\sigma_\text{NLO})/\sigma_\text{LO}$ is at the level of some $10^{-3}$.
\begin{figure}
    \centering
    \includegraphics[width=0.75\linewidth]{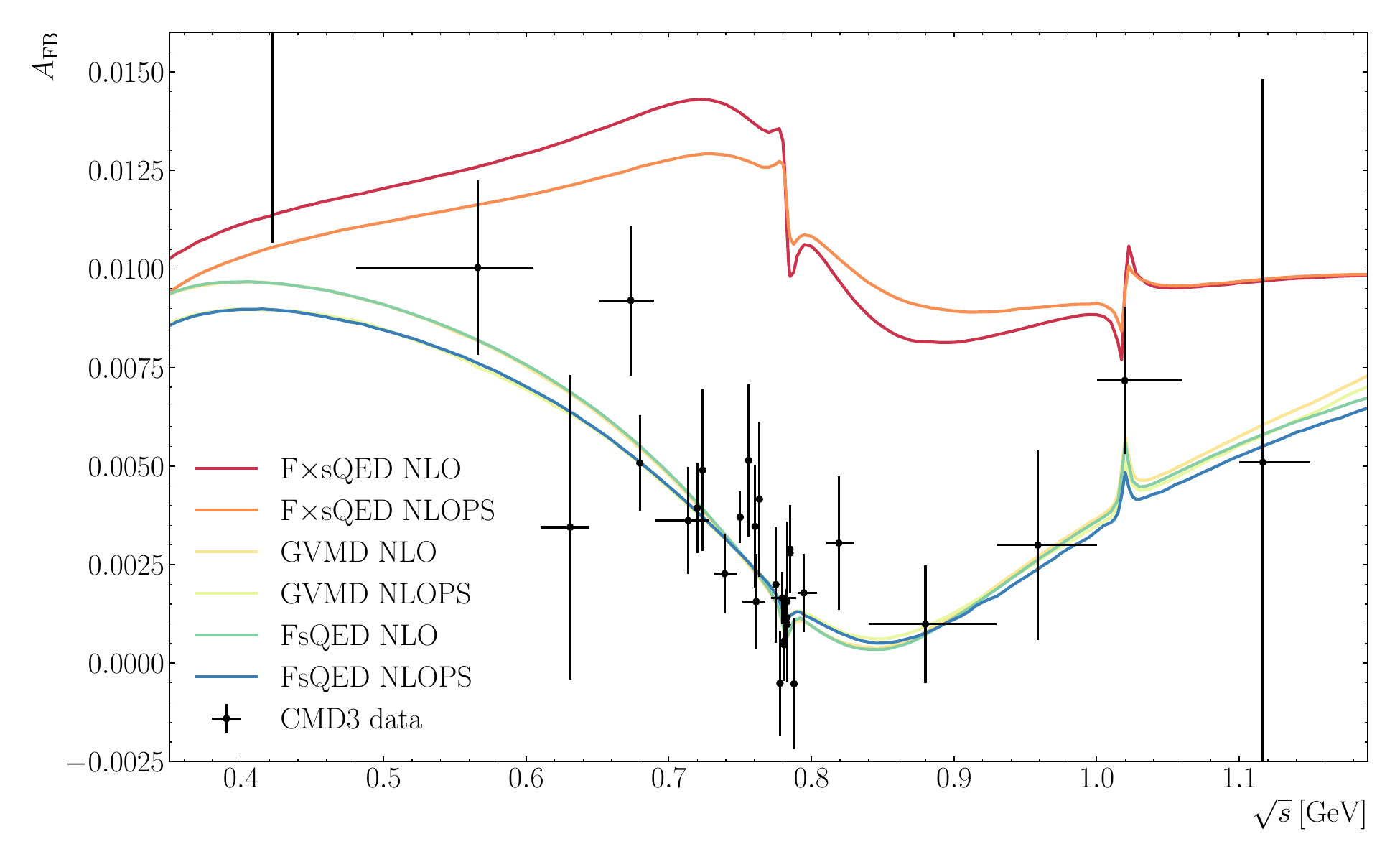}
    \caption{Forward-backward asymmetry as a function of the centre of mass energy with the F$\times$sQED, GVMD and FsQED approaches at two levels of precision. Black crosses are the experimental CMD3 data.}
    \label{fig:ASY}
\end{figure}
Finally, in fig.~\ref{fig:ASY} we show $A_\text{FB}(\sqrt{s})$ for the three approaches to $F_\pi(s)$ considered. As anticipated, the factorised approach is far from being a good description of data. On the other hand, the GVMD and FsQED approaches -- despite believed to be discrepant in the literature -- give the same picture, which is consistent with data. The effect of the PS on this observable is at the level of some per-cent, given the smallness of the value of the forward-backward asymmetry itself. 
\section{Conclusions}
Given the importance of the $\pi^+\pi^-$ production in $e^+e^-$ collisions at low energy, a precise theoretical description must take into account NLO corrections, incorporating both the pion form factor and higher order effects. We discussed briefly the available models that allow for the inclusion of $F_\pi(q^2)$ at loop-level and implemented it in an updated version of \textsc{BabaYaga@NLO}. Numerical results show that radiative corrections behave as expected and emphasize that properly accounting for the pion compositeness at NLO is necessary to describe loop-generated observables, such as the forward-backward asymmetry.
\acknowledgments The author thanks Ettore Budassi, Carlo M. Carloni Calame, Marco Ghilardi, Andrea Gurgone,  Guido Montagna, Mauro Moretti, Oreste Nicrosini and Fulvio Piccinini for fruitful collaboration.

\end{document}